\documentclass[aps,prb,twocolumn,amsmath,amssymb,groupedaddress, 
superscriptaddress,
longbibliography]{revtex4-2}

\usepackage{booktabs} 
\usepackage{graphicx}
\usepackage{dcolumn}
\usepackage{bm}
\usepackage{hyperref}

\usepackage{diagbox}

\begin{document}
\title{
Spiral-induced anomalous Hall effect from odd-parity spin-nodal lines
}
\author{Shun~Okumura} 
\email{okumura@ap.t.u-tokyo.ac.jp}
\affiliation{Quantum-Phase Electronics Center (QPEC), The University of Tokyo, Bunkyo, Tokyo 113-8656, Japan}
\affiliation{RIKEN Center for Emergent Matter Science (CEMS), Wako, Saitama 351-0198, Japan}
\affiliation{International Institute for Sustainability with Knotted Chiral Meta Matter (WPI-SKCM$^2$), Hiroshima University, Higashi-Hiroshima, Hiroshima 739-8531, Japan}
\author{Moritz~M.~Hirschmann}
\affiliation{RIKEN Center for Emergent Matter Science (CEMS), Wako, Saitama 351-0198, Japan}
\author{Yukitoshi~Motome}
\affiliation{Department of Applied Physics, The University of Tokyo, Bunkyo, Tokyo 113-8656, Japan}

\begin{abstract}
Spin spirals represent a fundamental class of noncollinear yet coplanar magnetic structures that give rise to diverse emergent phenomena through their coupling to itinerant electrons.
We investigate metallic systems with spin spirals and uncover an unconventional anomalous Hall effect (AHE) induced by spiral magnetism in both noncentrosymmetric and centrosymmetric crystals. 
The spin spiral generates odd-parity spin splitting with polarization perpendicular to the helical plane, forming spin-nodal lines in the electronic structure.
In the presence of spin--orbit coupling, finite magnetization opens gaps at these nodal lines, resulting in the Berry curvature concentrated near the gaps and a distinctive AHE. 
We identify the interplay among the spin--orbit coupling, helical plane orientation, and magnetization direction as the key ingredient for this spiral-induced AHE.
Our results establish a broadly applicable mechanism for anomalous Hall responses across a wide range of materials hosting spin spirals.
\end{abstract}

\maketitle

Off-diagonal transport phenomena in solids, particularly various Hall responses, constitute a central theme in condensed matter physics.
The conventional Hall effect arises from the Lorentz force exerted on charge carriers under an external magnetic field~\cite{Hall1879}, which explicitly breaks time-reversal symmetry.
In contrast, the anomalous Hall effect (AHE) in ferromagnets occurs even in the absence of a magnetic field~\cite{Hall1881}.  
This originates from Berry curvature~\cite{Karplus1954}, skew scattering~\cite{Smit1955, Smit1958}, and side-jump mechanisms~\cite{Berger1970} due to the synergy of net magnetization and spin--orbit coupling (SOC), which exemplifies the quantum interplay between spin, charge, and topology~\cite{Nagaosa2010}. 
Beyond these relativistic mechanisms, the discovery of the topological Hall effect (THE) in noncoplanar magnets~\cite{Taguchi2001} highlights an alternative origin based on scalar spin chirality, interpreted as a real-space Berry curvature~\cite{Ye1999, Ohgushi2000}. 
These developments have spurred intense research on topological transport phenomena in systems with nontrivial spin textures, such as magnetic skyrmions~\cite{Neubauer2009} and hedgehogs~\cite{Kanazawa2011}.

In recent years, antiferromagnets have emerged as a fertile platform for unconventional Hall responses. 
Despite having no net magnetization, many antiferromagnets break time-reversal symmetry and exhibit significant Hall responses. 
For instance, THE without net magnetization was proposed for all-in-all-out spin structures~\cite{Shindou2001, Martin2008, Akagi2010} and observed in pyrochlore iridates~\cite{Fujita2015, Ueda2018} and intercalated transition-metal dichalcogenides~\cite{Takagi2023, Park2023}. 
A coplanar 120$^\circ$ antiferromagnetic structure can also give rise to a remarkably large AHE~\cite{Chen2014, Nakatsuji2015}, attributed to the interplay between SOC and magnetic octupoles~\cite{Suzuki2017}. 
Furthermore, collinear antiferromagnets that lack bond inversion symmetry due to parity-breaking environments, referred to as altermagnets~\cite{Smejkal2022, Smejkal2022_2}, have recently attracted attention as a new class of AHE hosts, where SOC again plays a pivotal role~\cite{Solovyev1997, Smejkal2020}. 

\begin{figure}[b]
\centering
\includegraphics[width=\linewidth,clip]{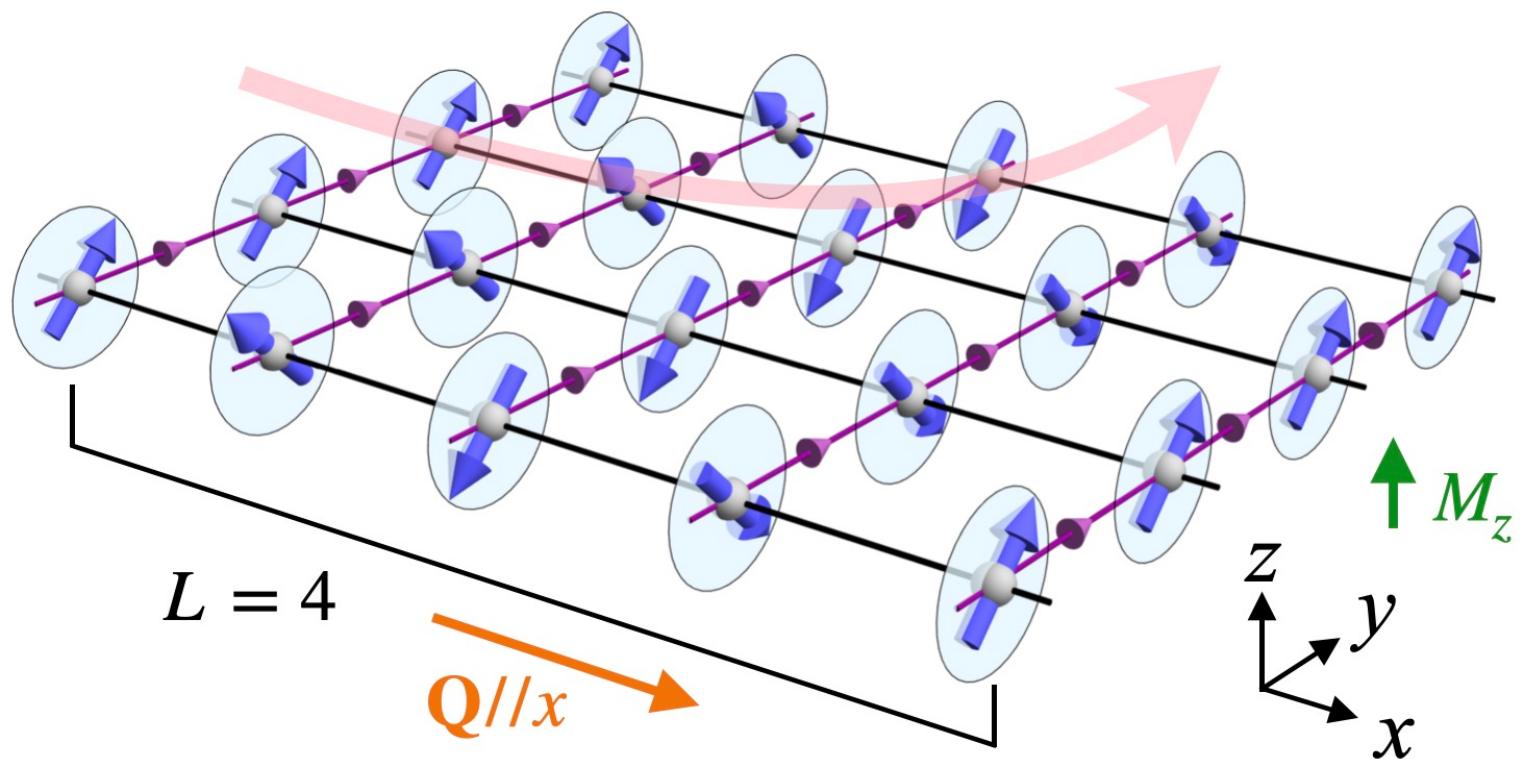}
\caption{
Schematic picture of the spiral-induced AHE.
The red arrow illustrates electron flow on a spin spiral order with the $yz$ helical plane and spiral pitch $L = 4$.
The blue arrows and circles on the square lattice denote localized spins and their helical planes, respectively.
The purple arrows denote the antisymmetric SOC on the $y$ bonds.
The green and orange arrows represent the net magnetization $M_z$ and the spiral propagation vector $\bold{Q}$ along the $x$ axis, respectively.
}
\label{f1}
\end{figure}

\begin{figure}[b]
\centering
\includegraphics[width=\columnwidth,clip]{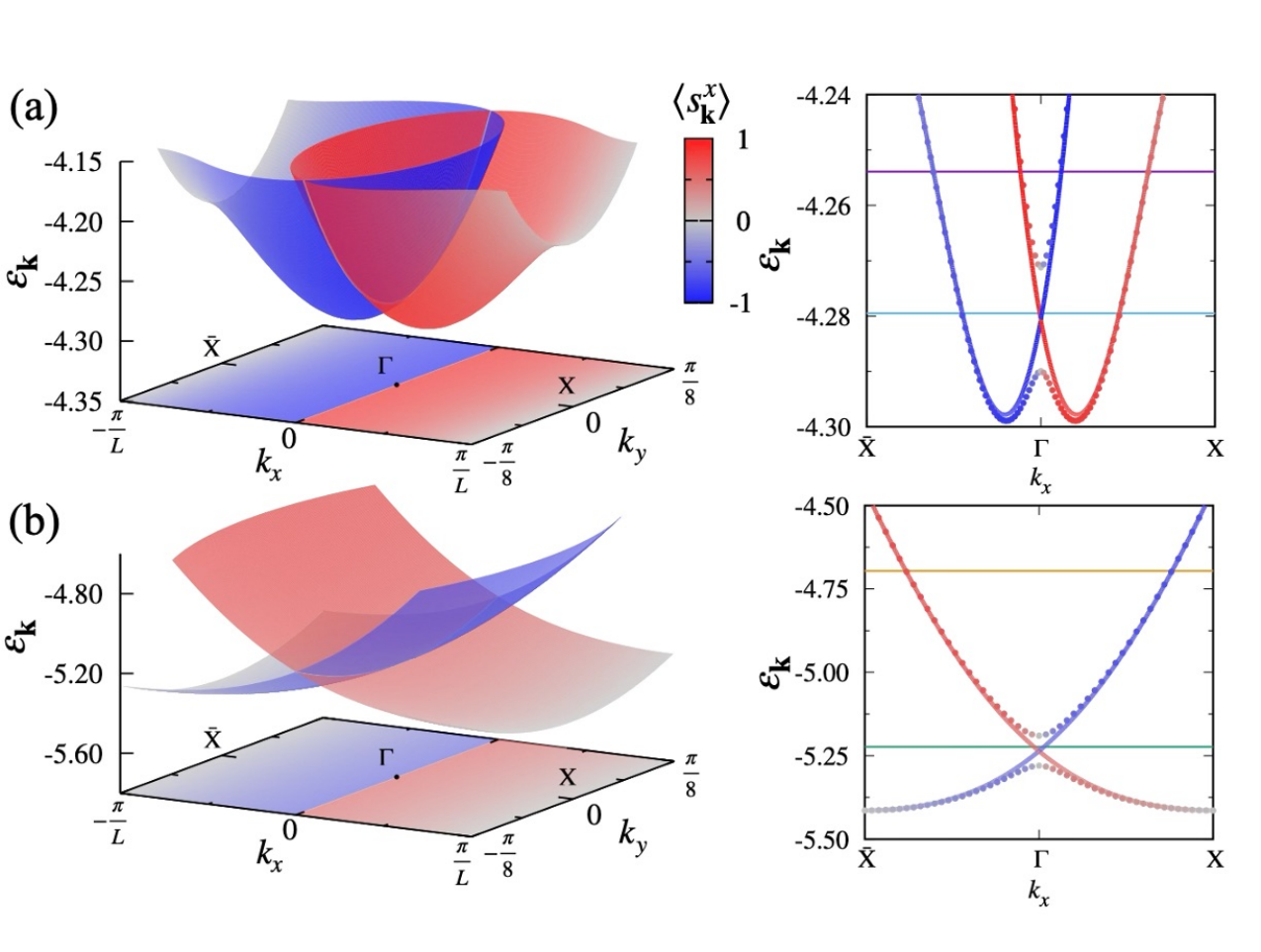}
\caption{
Odd-parity spin splitting and spin-nodal lines in low-energy electronic band structures for $\lambda=m=0$ in the $yz$ spiral state at (a) $J = 0.8$ and (b) $J = 2.0$.
The color represents the expectation value of the spin polarization along the  $x$ direction, $\langle s^x_{\bold{k}}\rangle$.
In the left panels, the lowest-energy bands are projected onto the bottom plane.
The right panels plot the energy dispersions along the $\bar{\mathrm{X}}\Gamma\mathrm{X}$ line at $k_y = 0$. 
The dotted lines show the results with SOC ($\nu=y$, $\lambda = 0.01$) and magnetization [$\mu=z$, $m = 0.03$ in (a), $m = 0.10$ in (b)]. 
The horizontal lines represent the Fermi levels $\varepsilon_{\mathrm{F}}$ for each electron filling $n_\mathrm{e}$ used for the AHC calculations in Fig.~\ref{f3}(a) and the Fermi surfaces in Figs.~\ref{f3}(b) and \ref{f3}(c).
}
\label{f2}
\end{figure}

By contrast, in spiral magnets, hosting noncollinear yet coplanar spin textures, their Hall response remains largely unexplored because they do not have a scalar spin chirality as a source of the THE. 
Spin spirals nevertheless represent a common class of compensated magnets, often stabilized in chiral~\cite{Dzyaloshinsky1958, Moriya1960}, frustrated~\cite{Yoshimori1959}, and itinerant magnets~\cite{Ruderman1954, Kasuya1956, Yosida1957}.
In particular, commensurate spin spirals have recently attracted growing interest as a platform for ``$p$-wave magnetism" characterized by nonrelativistic odd-parity spin splitting~\cite{Hellenes2023, Brekke2024, Chakraborty2025, Yu2025}, for which a distinctive AHE has been reported experimentally~\cite{Yamada2025}.
However, the microscopic origin of the Hall response in single-$Q$ spin spirals remains unclear, as prior studies focused mainly on more complex multiple-$Q$ spin textures.

In this Letter, we theoretically demonstrate an AHE in spin--charge coupled systems with spin spirals.
Our numerical calculations reveal that this spiral-induced anomalous Hall effect (SiAHE) stems from Berry curvature sharply concentrated near gapped spin-nodal lines created by odd-parity spin splitting.
The SiAHE exhibits a characteristic nonmonotonic dependence on magnetization, thereby bridging momentum-space Berry curvature and the underlying real-space spin texture.
A complementary low-energy model identifies the SOC type and magnetization direction required for a finite SiAHE. 
The sign of SiAHE is highly sensitive to the helical plane orientation, making the effect a powerful electrical probe for resolving spiral spin textures. 
We further demonstrate that this mechanism remains operative in centrosymmetric crystals, where inversion symmetry breaking by the spiral order allows spin splitting analogous to the antisymmetric SOC in noncentrosymmetric systems.

We investigate Hall responses in a two-dimensional electronic system that mimics an itinerant magnet, where electron motion is affected by coupling to a spin spiral texture.
We consider a spin--charge coupled model with SOC on a square lattice, as illustrated in Fig.~\ref{f1}, whose Hamiltonian is given by
\begin{eqnarray}
	\mathcal{H} &=& -t \sum_{\langle lj \rangle s}c^\dagger_{l s}c^{\;}_{j s} 
	+ i\sum_{\langle lj \rangle s s' }\bold{g}^{\;}_{l,j}\cdot c^\dagger_{l s}\boldsymbol{\sigma}^{\;}_{s s'}c^{\;}_{j s'} + {\rm h.c.} \nonumber\\
	&& -J \sum_{l s s'}c^\dagger_{l s}\boldsymbol{\sigma}^{\;}_{s s'}c^{\;}_{l s'}\cdot\bold{S}^{\;}_{l},
	\label{eq:lattice_model}	
\end{eqnarray}
where $c^{\;}_{ls}(c^{\dagger}_{ls})$ is an annihilation (creation) operator for an $s$-spin electron at site $l$ ($s = \uparrow$ or $\downarrow$), $\boldsymbol{\sigma} = (\sigma^x,\sigma^y,\sigma^z)$ is the vector of Pauli matrices, and ${\mathbf S}^{\;}_{l}$ is a localized spin at site $l$, treated as a classical vector with fixed length $\left|\bold{S}^{\;}_{l}\right|=1$.
The first term describes nearest-neighbor hopping of itinerant electrons with the transfer integral $t$.  
The second term denotes antisymmetric SOC characterized by $g$-vector $\bold{g}^{\;}_{l,j} = -\bold{g}^{\;}_{j,l}$ with amplitude $\lambda$.
To eliminate contributions from the conventional AHE due to SOC, we assume the SOC only on the $y$ bonds, namely, $\bold{g}_{l,l+\hat{\bold{e}}_y} = \lambda\hat{\bold{e}}_\nu$, otherwise zero.
The third term represents the onsite coupling between itinerant electrons and localized spins with strength $J$.
The spin--charge coupling is known to stabilize short-pitch spin spirals via the Ruderman–Kittel–Kasuya–Yosida~\cite{Ruderman1954, Kasuya1956, Yosida1957} or double-exchange mechanisms~\cite{Mostovoy2005, Azhar2017}. 
In this work, however, we prescribe the spin configuration to investigate the Hall response independently of its microscopic stabilization mechanism.

The spin spiral is characterized by a propagation vector $\bold{Q}=(Q,0,0)$ along the $x$ direction, where $Q = \frac{2\pi}{L}$ for a spiral pitch $L$.
The spin configuration is given as 
\begin{eqnarray}
	\bold{S}^{\;}_{l} = \bold{S}^{\alpha\beta;\mu}_{l} \propto \hat{\bold{e}}_{\alpha}\cos{\mathcal{Q}_l}
+ \hat{\bold{e}}_{\beta}\sin{\mathcal{Q}_l} + m \hat{\bold{e}}_{\mu},	
\end{eqnarray}
where $\alpha\beta$ denotes the helical plane and $\mu$ represents the magnetization direction; $\hat{\bold{e}}_\alpha$ is the unit vector along the $\alpha$ direction; and $\mathcal{Q}_l = \bold{Q}\cdot\bold{r}_l + \varphi$ with position vector $\bold{r}_l$ and phase $\varphi$. 
The parameter $m$ controls the uniform magnetization superposed on the spin spiral.
For generality, its microscopic origin is not specified in the present model; it may arise either from an external magnetic field or intrinsically from incomplete compensation of the spin spiral coupling to the electronic structure~\cite{Okumura2018}.
 The net magnetization $M_\mu$ is calculated as $M_\mu = \frac{1}{N}\sum_{l}\bold{S}^{\;}_{l}\cdot\hat{\bold{e}}_\mu$, where $N$ is the total number of sites.
In the following calculations, we set $t = 1$ as an energy unit, $\lambda = 0.01$, and $N = 200^2 L$.
Unless otherwise stated, we take $L = 4$ and $\varphi  = \frac{\pi}{L}$ (see Fig.~\ref{f1}).
Further computational details are described in the Supplemental Material~\cite{SM}.

We begin by examining the electronic band structure on the spin spiral. 
Figure~\ref{f2} shows the low-energy bands for the $yz$ spiral state illustrated in Fig.~\ref{f1} (the whole band structures are shown in Fig.~\ref{f4}). 
For small $J$, the parabolic bands exhibit a slight shift along the propagation direction of the spin spiral, $k_x$, [Fig.~\ref{f2}(a)], while for large $J$, the band splitting becomes pronounced and the band bottom reaches the first Brillouin zone (BZ) edge [Fig.~\ref{f2}(b)].
Each shifted band carries spin polarization $\langle s^x_{\bold{k}} \rangle$ perpendicular to the $yz$ helical plane, whose sign is determined by the spiral chirality.
This chirality-dependent spin polarization explicitly reflects the breaking of space inversion symmetry by the spiral order.
Such odd-parity band splitting, accompanied by spin-nodal lines, is characteristic of the recently proposed $p$-wave magnet with even-period commensurate spirals~\cite{Hellenes2023, Brekke2024}. 
It is, however, a general property of spin spirals with arbitrary periods and can be explained by a U(1) gauge transformation~\cite{Okumura2018, Okumura2021}.

In the absence of SOC, the electronic band structure exhibits a spin-unpolarized nodal line along the intersections of the spin-split bands, as shown in Fig.~\ref{f2}.
These spin-nodal lines are protected by spin symmetries $C^{x}_{\pi}\mathcal{T}$ and $C^{x}_{Q} T^{\;}_{x}$, comprising a $\theta$-spin rotation around the $x$ axis perpendicular to the $yz$ helical plane, $C^{x}_{\theta}$, a time-reversal operation, $\mathcal{T}$, and a one-site translation in the $x$ direction, $T^{\;}_{x}$.
The operation $C^{x}_{\pi}\mathcal{T}$ enforces $\langle{s^{x}_\bold{k}}\rangle = -\langle{s^{x}_\bold{-k}}\rangle$, and $C^{x}_{Q}T^{\;}_{x}$ implies $\langle{s^{y}_\bold{k}}\rangle = \langle{s^{z}_\bold{k}}\rangle = 0$ for each band.
These spin symmetries ensure the appearance of the spin-nodal lines at $k_x = 0$ and $\pm\frac{\pi}{L}$, a defining feature of odd-parity spin splitting in spin spirals~\cite{Hellenes2023, Yamada2025}. 
  
Both symmetries are generally broken by SOC and uniform magnetization.
We first consider antisymmetric SOC with uniform $g$-vectors and discuss the centrosymmetric case later.
The dotted lines in the right panels of Figs.~\ref{f2}(a) and \ref{f2}(b) show the band structures in the presence of SOC $\lambda$ ($\nu = y$) and magnetization $m$ ($\mu = z$), under which the spin spiral develops a solitonic modulation within the helical plane.
In this situation, time-reversal symmetry is broken, and the spin-nodal lines become gapped, while the $k_x$-dependent spin polarization is preserved. 
The gap opening at the spin-nodal lines gives rise to an anomalous Hall response, which we call the SiAHE. 
In the present setup, where antisymmetric SOC is restricted to the $y$ bonds, mirror symmetry forbids the conventional AHE in the absence of the spin spiral, even at finite net magnetization. 
Including SOC on the $x$ bonds yields only a negligible additional contribution to the SiAHE
(see Supplemental Material~\cite{SM}).

\begin{figure}[t]
\centering
\includegraphics[width=\columnwidth,clip]{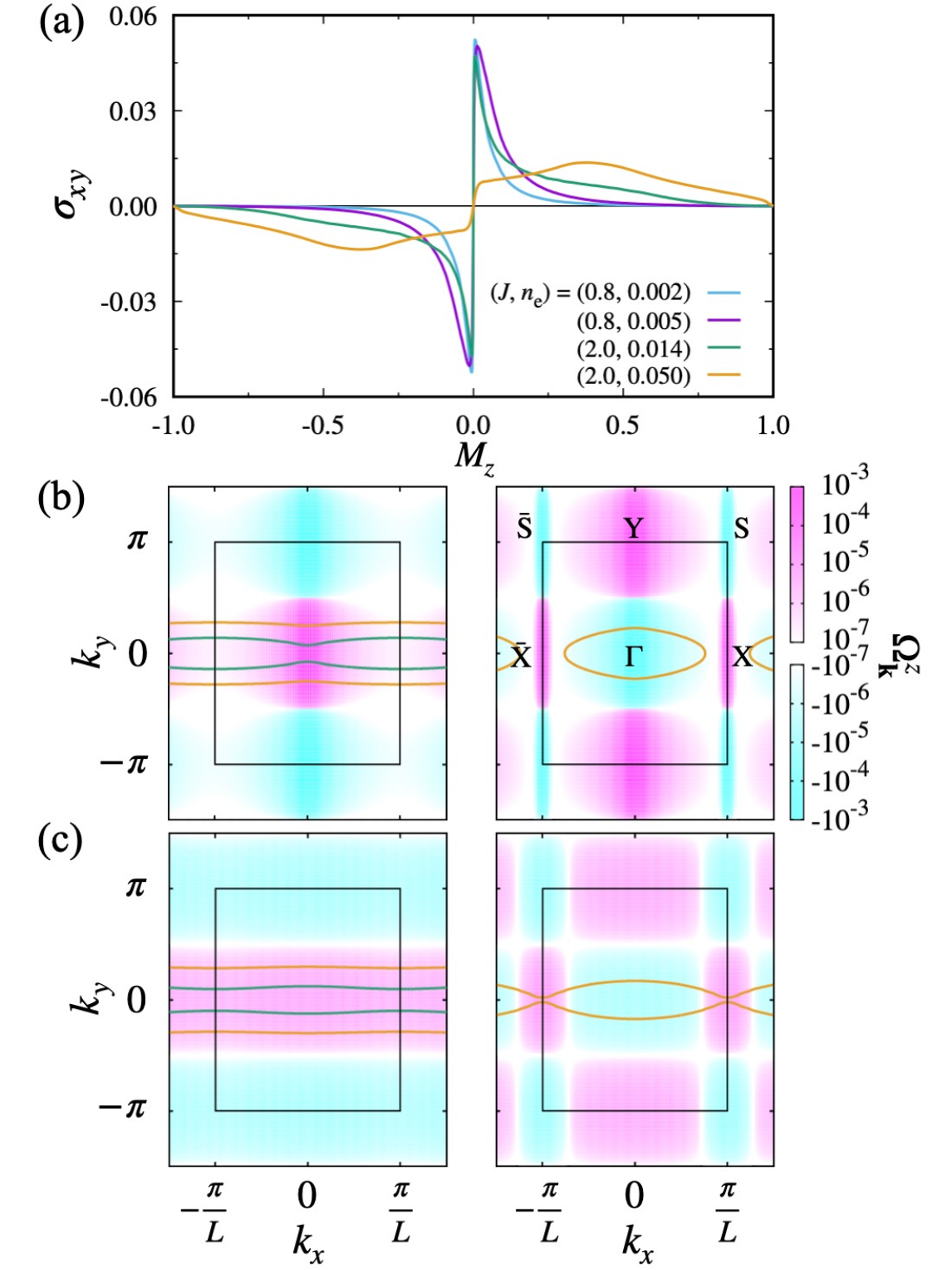}
\caption{
(a) AHC $\sigma_{xy}$ as a function of the net magnetization $M_z$ in the $yz$ spiral state for several parameter sets of $J$ and $n_\mathrm{e}$. 
(b), (c) Color plots of the Berry curvature $\Omega^z_{\bold{k}}$ of the lowest (left) and the second-lowest (right) energy bands for $J = 2.0$ at (b) $m = 0.1$ ($M_z \sim 0.05$) and (c) $m = 0.5$ ($M_z \sim 0.3$).
The black rectangles denote the first BZ. 
The green and orange lines represent the Fermi surfaces at $n_\mathrm{e} = 0.014$ and $0.050$, respectively.
}
\label{f3}
\end{figure}

Figure~\ref{f3}(a) shows the anomalous Hall conductivity (AHC) $\sigma_{xy}$ as a function of the net magnetization $M_z$ for $J = 0.8$ and $2.0$ at several electron densities $n_\mathrm{e}$.
Here, the Fermi energy $\varepsilon_\mathrm{F}$ is adjusted to keep $n_\mathrm{e}$ fixed while varying $M_z$.
In all cases, $\sigma_{xy}$ is an odd function of $M_z$ and exhibits a nonlinear dependence with a sharp increase near $M_z = 0$.
This characteristic behavior originates from Berry curvature concentrated near the gapped spin-nodal lines.
Figures~\ref{f3}(b) and \ref{f3}(c) show the Berry curvature distributions $\Omega^{z}_\bold{k}$ in the two lowest-energy bands at $J = 2.0$ for $m = 0.1$ and $0.5$, respectively.
These plots reveal pronounced Berry curvature along the $\Gamma$-Y line in both bands, as well as along the X-S line in the second-lowest energy band, where the spin expectation value vanishes at $m = 0$.
Note that the Berry curvature obeys $\Omega^{z}(k_x,k_y) = \Omega^{z}(-k_x,k_y)$ as enforced by $C_{2x}\mathcal{T}$ symmetry. 
In addition, for the spiral phase $\varphi = \frac{\pi}{L}$ adopted in the present calculations, the Hamiltonian satisfies spinful $C_{2z}$ symmetry, $\sigma_z H_\bold{k}\sigma_z = H_{-\bold{k}}$, which leads to $\Omega^{z}(k_x, k_y) = \Omega^{z}(-k_x,-k_y)$. 
While the latter does not hold for arbitrary $\varphi$, the $C_{2x}\mathcal{T}$ constraint remains intact for general spin spirals.

While the Berry curvature integrates to zero within each band, resulting in all bands being topologically trivial with vanishing Chern numbers, a finite AHC emerges in the metallic regime.
When the Fermi energy $\varepsilon_\mathrm{F}$ is located in the vicinity of the spin-nodal line at the $\Gamma$ point, the AHC decreases rapidly with increasing $M_z$ for both $J = 0.8$ and $2.0$, as shown by the light-blue and green lines in Fig.~\ref{f3}(a), respectively. 
In this regime, the SiAHE is dominated by contributions from the lowest-energy band and depends only weakly on $J$.
For example, at $J = 2.0$ and $m=0.1$ ($M_z \sim 0.05$), $\sigma_{xy}$ mainly originates from the positive $\Omega^{z}_\bold{k}$ around the $\Gamma$ point in the lowest-energy band [Fig.~\ref{f3}(b)]. 
At $m=0.5$ ($M_z \sim 0.3$), however, it is suppressed since $\Omega^{z}_\bold{k}$ becomes broadly distributed throughout the BZ, accompanied by a larger energy gap.

For more general $\varepsilon_\mathrm{F}$, the behavior differs markedly between the weak- and strong-$J$ cases.
For $J = 0.8$, $\sigma_{xy}$ for $n_\mathrm{e} = 0.005$, where $\varepsilon_\mathrm{F}$ lies well above the nodal line [Fig.~\ref{f2}(a), right], behaves similarly to that for $n_\mathrm{e} = 0.002$, where $\varepsilon_\mathrm{F}$ is close to the nodal line.
For $J = 2.0$, however, $\sigma_{xy}$ for $n_\mathrm{e} = 0.050$ exhibits a broad peak at $M_z \sim 0.4$ and remains large up to saturation $M_z = 1$.
When $J$ is large, as shown in Figs.~\ref{f3}(b) and \ref{f3}(c), the Fermi surfaces become highly anisotropic due to the large band splitting and take distinct shapes in the lowest and second-lowest energy bands.
Moreover, an additional contribution to $\sigma_{xy}$ arises from the gapped spin-nodal line along the X-S line at large $M_z$ as shown in Fig.~\ref{f3}(c).
These contribute to the qualitatively different behavior of $\sigma_{xy}$ across a wide field range.

\begin{figure}[t]
\centering
\includegraphics[width=\columnwidth,clip]{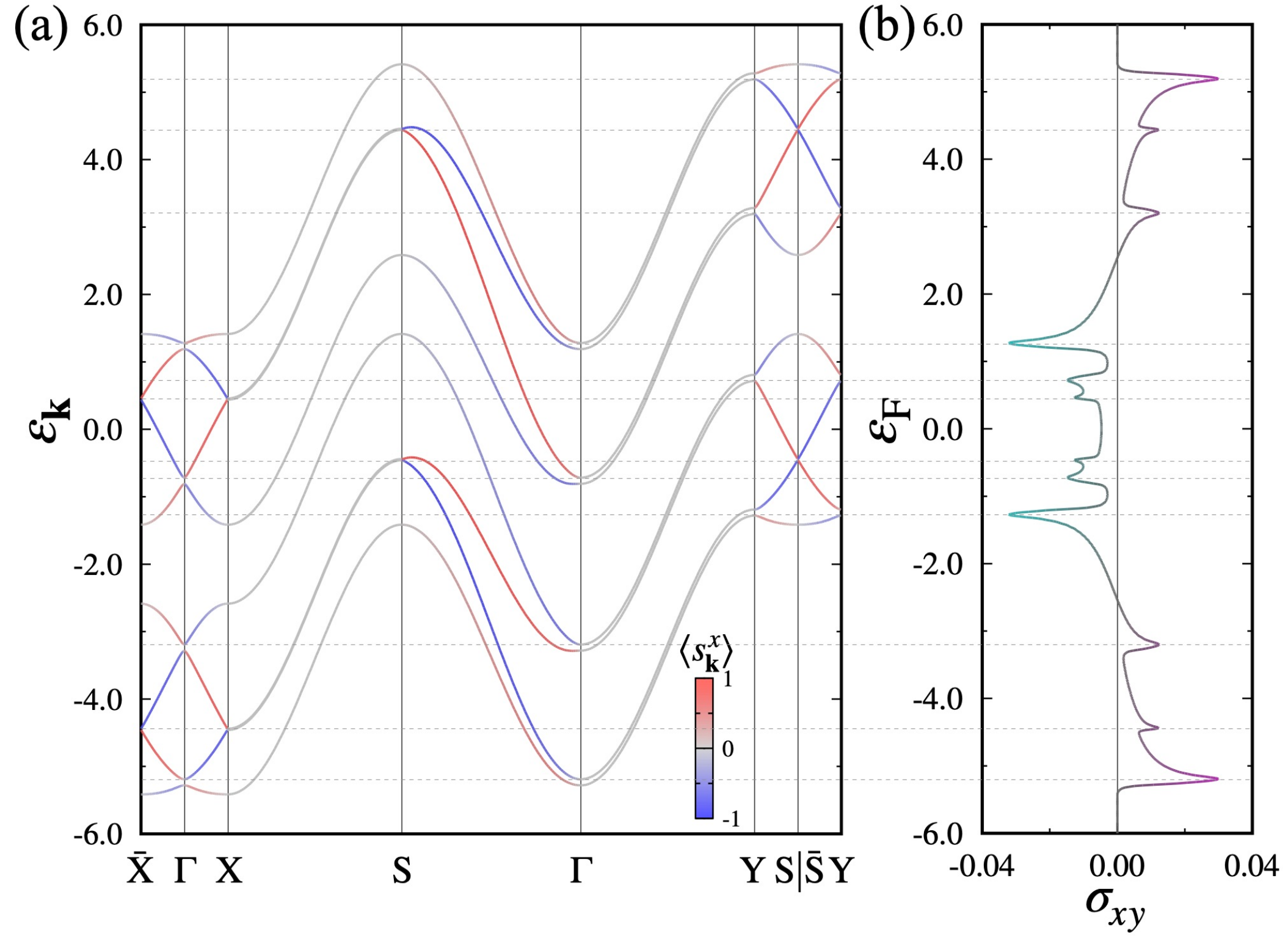}
\caption{
(a) Full electronic band structure along the representative symmetric lines in the first BZ, colored by the spin polarization $\langle s^{x}_{\bold{k}}\rangle$, at $J=2.0$, $\lambda=0.01$, and $m = 0.1$ ($M_z \sim 0.05$).  
The high-symmetry points are denoted in Fig.~\ref{f3}(b). 
(b) Corresponding $\sigma_{xy}$ in the $yz$ spiral state as a function of $\varepsilon_\mathrm{F}$.
The horizontal dashed lines represent the energy levels where $\sigma_{xy}$ has peaks, as guides for the eyes.
}
\label{f4}
\end{figure}

Our findings indicate that the SiAHE is deeply linked to the emergence of spin-nodal lines and anisotropic Fermi surfaces arising from odd-parity spin splitting under spin spirals.
To examine this relationship beyond the low-energy regime, we systematically investigate the Fermi energy dependence of the AHC.
As shown in Fig.~\ref{f4}(a), the full band structure contains twelve spin-nodal lines along the $\Gamma$-Y and X-S lines. 
All these nodal lines open gaps under finite $M_z$, generating Berry curvature that drives SiAHE.
Figure~\ref{f4}(b) shows $\sigma_{xy}$ as a function of $\varepsilon_\mathrm{F}$ for $J = 2.0$.
Each peak of $\sigma_{xy}$ corresponds to the energy of a spin-nodal line.
This demonstrates that the SiAHE caused by spin-nodal lines is not limited to the low-energy regime, but is ubiquitously observed across a wide range of electron fillings.

\begin{table}[t]
\caption{
Conditions for the SiAHE for a given helical plane and corresponding spiral-induced spin polarization.
Notations of $\nu$ and $\mu$ follow the effective model.
The symbols $\pm$ represent a nonzero AHC with opposite signs, while a dash indicates a vanishing AHC.
}
\label{t1}
\centering
\renewcommand\arraystretch{1.4}
\begin{tabular}{c c ccc c ccc c ccc c}
  \hline \hline
  helical plane $\alpha\beta$ & & & $xy$ & & & & $yz$ & & & & $zx$ & &\\
  \hline
  spin polarization & & & $\langle s^{z}_{\bold{k}}\rangle$ & & & & $\langle s^{x}_{\bold{k}}\rangle$ & & & & $\langle s^{y}_{\bold{k}}\rangle$ & & \\
  \hline
  \diagbox{$\nu$}{$\mu$} & & $x$ & $y$ & $z$ & \hspace{0.5cm} & $x$ & $y$ & $z$ & \hspace{0.5cm} & $x$ & $y$ & $z$ & \hspace{0.2cm} \\
  \hline
  $x$	&	& - & $\pm$ & - &	&  - & - & - &		& - & - & $\mp$	& \\
  $y$ &	& $\mp$ & - & - &	&  - & - & $\pm$ &	& - & - & - 		& \\
  $z$	&	& - & - & - &		&  - & $\mp$ & - &	& $\pm$ & - & -	& \\
  \hline \hline
\end{tabular}
\end{table}

Here, we discuss the necessary conditions for SiAHE, adopting a low-energy effective model in the continuum limit. 
In the vicinity of the $\Gamma$ point, the system is described by the two-level Hamiltonian as $H^{\mathrm{eff}}_{\bold{k}} \equiv d_{\bold{k}}^0\sigma^0 + \bold{d}_{\bold{k}}\cdot\boldsymbol{\sigma}$, where $\sigma^0$ is an identity matrix, $d^0_\bold{k} = t(k_x^2 + k_y^2)$, $d^\rho_\bold{k} = -pk_x$, $d^\nu_\bold{k} = \tilde{\lambda}k_y$, and $d^\mu_\bold{k} = -\tilde{m}$; $p$ is the spin polarization parameter, perturbatively obtained as $p = \frac{J^2}{2t}\frac{\sin Q}{(1-\cos Q)^2}$~\cite{Yamada2025}, and $\tilde{\lambda}$ and $\tilde{m}$ are reduced SOC and magnetization, respectively.
The Berry curvature in this low-energy model is expressed by 
\begin{eqnarray}
	\Omega^z_{\bold{k}\pm}=\mp\epsilon^{\rho\nu\mu}d_{\bold{k}}^\mu\partial_{k_x}d_{\bold{k}}^\rho\partial_{k_y}d_{\bold{k}}^\nu /(2|\bold{d}_\bold{k}|^3),
	\label{eq:Berry}	
\end{eqnarray}
for the upper ($+$) and lower ($-$) bands. 
Hence, nonzero $\sigma_{xy}$ requires $\rho$, $\nu$, and $\mu$ to represent three mutually orthogonal spin components.
Table~\ref{t1} summarizes the combinations of the SOC component and uniform-magnetization direction that satisfy this condition.
As the spiral-induced spin polarization is perpendicular to the helical plane, the remaining two components must be completed with SOC and uniform magnetization.
The expression for the Berry curvature $\Omega^z_{\bold{k}\pm}$ in Eq.~\eqref{eq:Berry} further shows that its sign is reversed when the spin components are swapped between SOC and magnetization.

\begin{figure}[t]
\centering
\includegraphics[width=\columnwidth,clip]{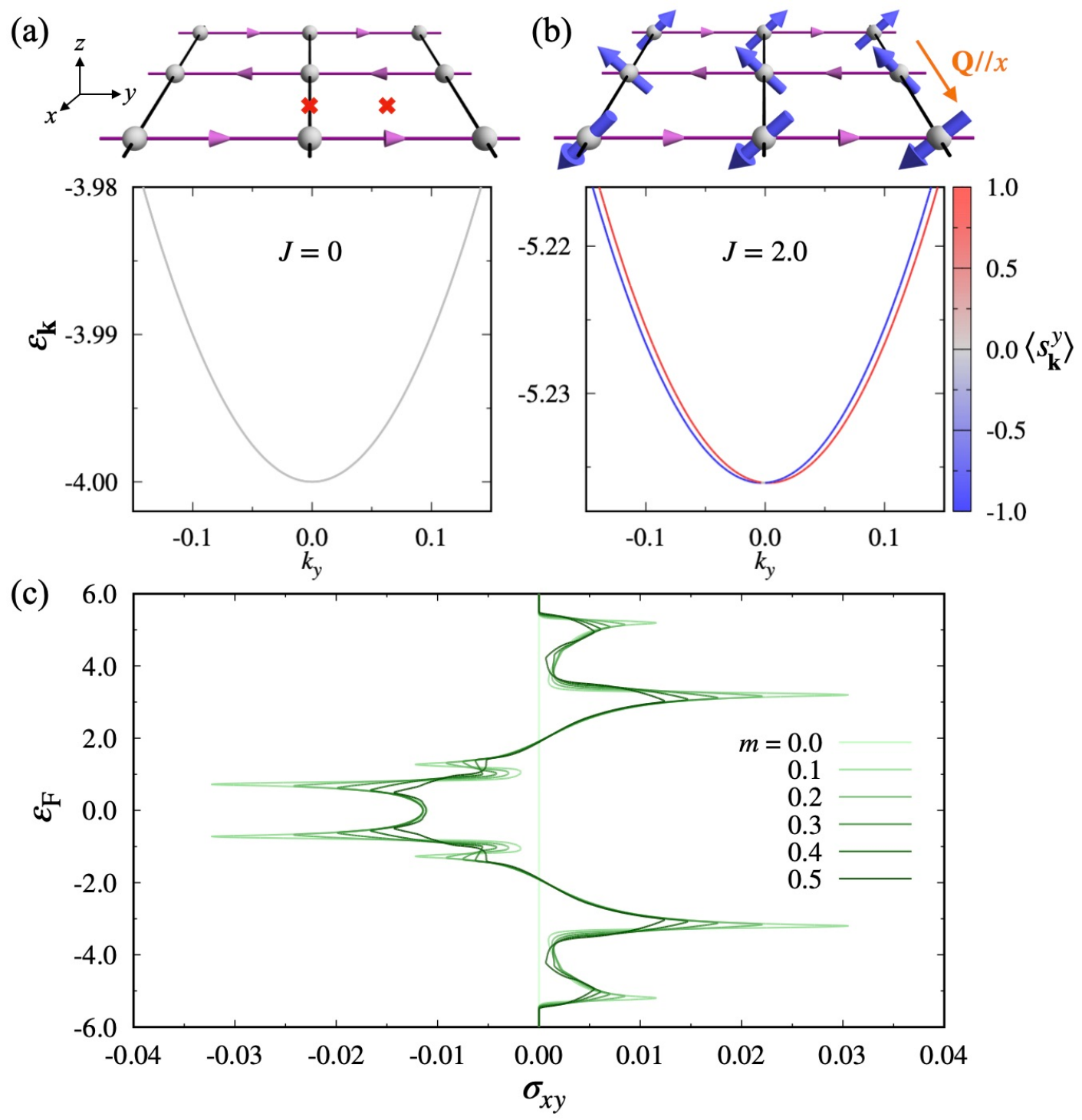}
\caption{
(a), (b) Schematics of the staggered-SOC model and low-energy band dispersions at $k_x = 0$, $m = 0$, and $\lambda = 0.01$, for (a) $J = 0$ and (b) $J=2.0$, colored by the spin polarization $\langle s^{y}_{\bold{k}}\rangle$.  
In (a), the inversion centers are marked by the red crosses.
(c) $\sigma_{xy}$ as a function of the Fermi energy $\varepsilon_{\mathrm F}$ for several values of $m$ in the $yz$-spiral state at $J=2.0$.
}
\label{f5}
\end{figure}

Thus far, we have considered noncentrosymmetric systems with antisymmetric SOC. 
We now show that the SiAHE can also arise in centrosymmetric crystals, where space inversion symmetry is broken by the spiral order rather than by the underlying crystal structure. 
To demonstrate this, we replace the uniform SOC with a staggered SOC whose sign alternates along the $x$ direction, $\bold{g}^{\;}_{l,l+\hat{\bold{e}}_y} = (-1)^{l_x}\lambda \hat{\bold{e}}_z$, where $l_x$ is the $x$ coordinate of site $l$, while retaining SOC only on the $y$ bonds. 
At $J = 0$, the Hamiltonian preserves both inversion and time-reversal symmetries, ensuring doubly degenerate bands even with finite $\lambda = 0.01$ [Fig.~\ref{f5}(a)]. 
For finite $J$, however, coupling to the $yz$-spin spiral breaks inversion symmetry and produces odd-parity spin splitting along $k_y$ with spin polarization in the $y$ direction, as shown in Fig.~\ref{f5}(b). 
This demonstrates that the interplay between the spiral order and staggered SOC generates an effective $k_y \sigma_y$ term analogous to that produced by uniform antisymmetric SOC, allowing the symmetry-based argument developed above to be applied. 
Consistent with this picture, a finite magnetization $m$ yields a nonzero AHC as plotted in Fig.~\ref{f5}(c). 
At the same SOC strength $\lambda = 0.01$, $\sigma_{xy}$ is comparable in magnitude to that obtained with uniform antisymmetric SOC in Fig.~\ref{f4}(b).
Similar to the uniform-SOC case, a pronounced magnetization dependence appears when $\varepsilon_\mathrm{F}$ lies near the gapped spin-nodal lines, whereas away from these energies, $\sigma_{xy}$ depends only weakly on $m$ over the range shown in Fig.~\ref{f5}(c).
These results establish that structural inversion asymmetry is not necessary for the SiAHE and support the applicability of the proposed mechanism to centrosymmetric spiral magnets.

We have theoretically demonstrated that spin spirals can induce a characteristic AHE via odd-parity spin splitting in the electronic band structure, under appropriate SOC and magnetization. 
Crucially, structural inversion asymmetry is not a prerequisite since spiral order can break inversion symmetry in a centrosymmetric crystal and, together with staggered SOC, enable the same Hall mechanism.
Our results provide a possible microscopic mechanism relevant to the AHE recently observed in a metallic $p$-wave magnet with a commensurate spin spiral and a centrosymmetric crystal structure~\cite{Yamada2025}. 
They also establish design principles applicable to a wide variety of spiral magnets.

The resulting SiAHE exhibits a characteristic nonmonotonic magnetization dependence despite arising from a simple single-$Q$ coplanar spin texture without scalar spin chirality, in contrast to complex multiple-$Q$ spin textures that generate THE.
It is also conceptually distinct from the previously proposed chiral Hall effect arising from higher-order gradients of real-space spin textures~\cite{Lux2020, Bouaziz2021} since the SiAHE originates purely from momentum-space Berry curvature.
Importantly, the SiAHE can manifest as a planar Hall response driven by in-plane magnetization~\cite{Ezawa2025arXiv_PHE}, in stark contrast to the conventional AHE of ferromagnets.

The emergence of SiAHE is highly sensitive to the helical plane orientation and the SOC type, yet remarkably insensitive to details such as the Fermi energy, spin-charge coupling, or even the parity of the spiral pitch (see Supplemental Material~\cite{SM}).
Our symmetry analysis indicates that this mechanism can operate not only in even-period commensurate spirals, but also in odd-period spirals and incommensurate spirals. 
Although odd-parity spin splitting need not occur in every band in these cases, the relevant spin symmetries remain operative at the low-energy band crossing at the $\Gamma$ point in the absence of SOC and net magnetization.
The spin splitting near this crossing can therefore be described by the same low-energy effective model developed above. 
SOC and magnetization satisfying the identified symmetry conditions can then gap the crossing and generate Berry curvature, enabling a finite SiAHE in the metallic regime. 
Thus, the SiAHE constitutes a general anomalous Hall mechanism in spiral magnets under appropriate SOC and magnetization conditions, rather than a phenomenon specific to even-period commensurate order.

Furthermore, in contrast to the AHE arising from a coplanar $120^\circ$ spin structure, which corresponds to ferroic cluster magnetic octupole order on a three-sublattice structure~\cite{Suzuki2017}, the SiAHE does not rely on a specific underlying lattice structure.
Together with its extension to centrosymmetric crystals, this property establishes the SiAHE as a broadly applicable mechanism for spiral magnets, with potential material platforms including rare-earth compounds, such as $R$Be$_{13}$~\cite{Bouree-Vigneron1991}.
Since the effect directly reflects the spiral chirality, topology of spin-nodal lines, and anisotropic spin splitting in the electronic structure, it offers a powerful electrical probe of spiral spin textures without relying on scattering experiments.
The present SiAHE is entirely intrinsic in origin, arising from the Berry curvature of the electronic bands, while possible extrinsic contributions associated with impurity scattering remain an intriguing direction for future investigation.
More broadly, our work deepens the insights into topological properties arising from odd-parity spin splitting~\cite{Martin2012, Ezawa2024, Nagae2025, Sukhachov2025, Bobkov2025}, and opens avenues toward electrical control in spiral magnets~\cite{Murakawa2009, Jiang2020, Masuda2024, Song2025} and nonlinear functionalities in noncollinear magnetic textures~\cite{Aoki2019, Okumura2021, Mayo2025, Sivianes2025, Nakamura2025, Deaconu2025, Ezawa2025, Ezawa2025_PVE}.

\begin{acknowledgments}
We would like to thank M.~Ezawa, M.~Hirschberger, J.~Masell, Y.~Onose, and R.~Yamada for fruitful discussions.
This research was supported by the JSPS KAKENHI (Nos.~JP22K13998, JP23K25816, JP25H01247, and JP26H00634) and JST PRESTO (No.~JPMJPR2595).
M.M.H. is funded by the Deutsche Forschungsgemeinschaft (DFG, German Research Foundation) - project No.~518238332 and by the RIKEN Special Postdoctoral Researcher Program.
Parts of the numerical calculations were performed on the supercomputing systems at ISSP, the University of Tokyo.
\end{acknowledgments}

\newpage
\section*{Supplemental Material for ``Spiral-induced anomalous Hall effect from odd-parity spin-nodal lines"} 
\subsection*{Calcuration details}

We present methods for calculating physical quantities in momentum space.
We perform a Fourier transform of the real-space Hamiltonian in Eq.~(1) in the main text into $\mathcal{H} = \sum_{\bold{k}}\vec{c}^{\;\dagger}_{\bold{k}}H^{\;}_{\bold{k}}\vec{c}^{\;}_{\bold{k}}$, where $H_\bold{k}$ is the $2L\times2L$ matrix and $\vec{c}_\bold{k}$ is a vector with $2L$ components.
By diagonalizing $H_{\bold{k}}$ for each wavenumber $\bold{k}$, we obtain the $a$th eigenenergy $\varepsilon_{\bold{k}a}$ and eigenfunction $|\bold{k}a\rangle$.
We calculate the spin polarization of the $a$th band as $\langle s^{\eta}_{\bold{k}a}\rangle = \langle\bold{k}a|\sigma^\eta\otimes I_L|\bold{k}a\rangle$; $I_L$ is an $L\times L$ identity matrix.

Using the linear response theory, we also calculate the anomalous Hall conductivity 

\begin{eqnarray}
	\sigma_{xy} = \frac{e^2}{\hbar}\int_\mathrm{BZ}d\bold{k}\sum_{a}\Omega^z_{\bold{k}a}f(\varepsilon_{\bold{k}a}),
	\label{eq:AHC}
\end{eqnarray} 
where $\Omega^z_{\bold{k}a}$ is the Berry curvature of the $a$th band given by 
\begin{eqnarray}
	\Omega^z_{\bold{k}a} = -i\epsilon^{\alpha\beta z}\sum_{b(\neq a)}\frac{\langle\bold{k}a|\partial_{k_\alpha}H_{\bold{k}}|\bold{k}b\rangle\langle\bold{k}b|\partial_{k_\beta}H_{\bold{k}}|\bold{k}a\rangle}{(\varepsilon_{\bold{k}a} - \varepsilon_{\bold{k}b})^2},
	\label{eq:Berry_curvature}
\end{eqnarray}
$\epsilon^{\alpha\beta\gamma}$ is the Levi-Civita, and $f(\varepsilon_{\bold{k}a}) = \frac{1}{e^{(\varepsilon_{\bold{k}a} - \varepsilon_\mathrm{F})/T}+1}$ is the Fermi-Dirac distribution function at temperature $T$.
The Fermi energy $\varepsilon_\mathrm{F}$ is determined by the given electron density $n_\mathrm{e} = \frac{1}{N}\sum_{\bold{k}a}f(\varepsilon_{\bold{k}a})$, and the integral is taken in the first Brillouin zone (BZ).
In the main text, we fix the temperature $T = 0.01t$ and take the elementary charge $e=1$, the reduced Planck constant $\hbar=1$, and the lattice constant $a_0=1$.

\subsection*{Effect of the $x$-bond spin--orbit coupling}

In the main text, we considered antisymmetric SOC only on the $y$ bonds to eliminate the conventional AHE arising solely from SOC. 
In realistic materials, however, SOC is generally expected to be present on both the $x$ and $y$ bonds. 
To examine the robustness of our results, we additionally introduce SOC on the $x$ bonds, $ \mathbf{g}_{l,l+\hat{x}}=\lambda_x \hat{\mathbf{e}}_x$, and compare the resulting AHC for $\lambda_x = \pm 0.01$ in the same setup as in Fig.~\ref{f3} of the main text.
For the ferromagnetic reference, we replace the spiral exchange term by a spatially uniform exchange field $-JM_z\sigma^z$.

\begin{figure}[t]
\centering
\includegraphics[width=0.8\columnwidth,clip]{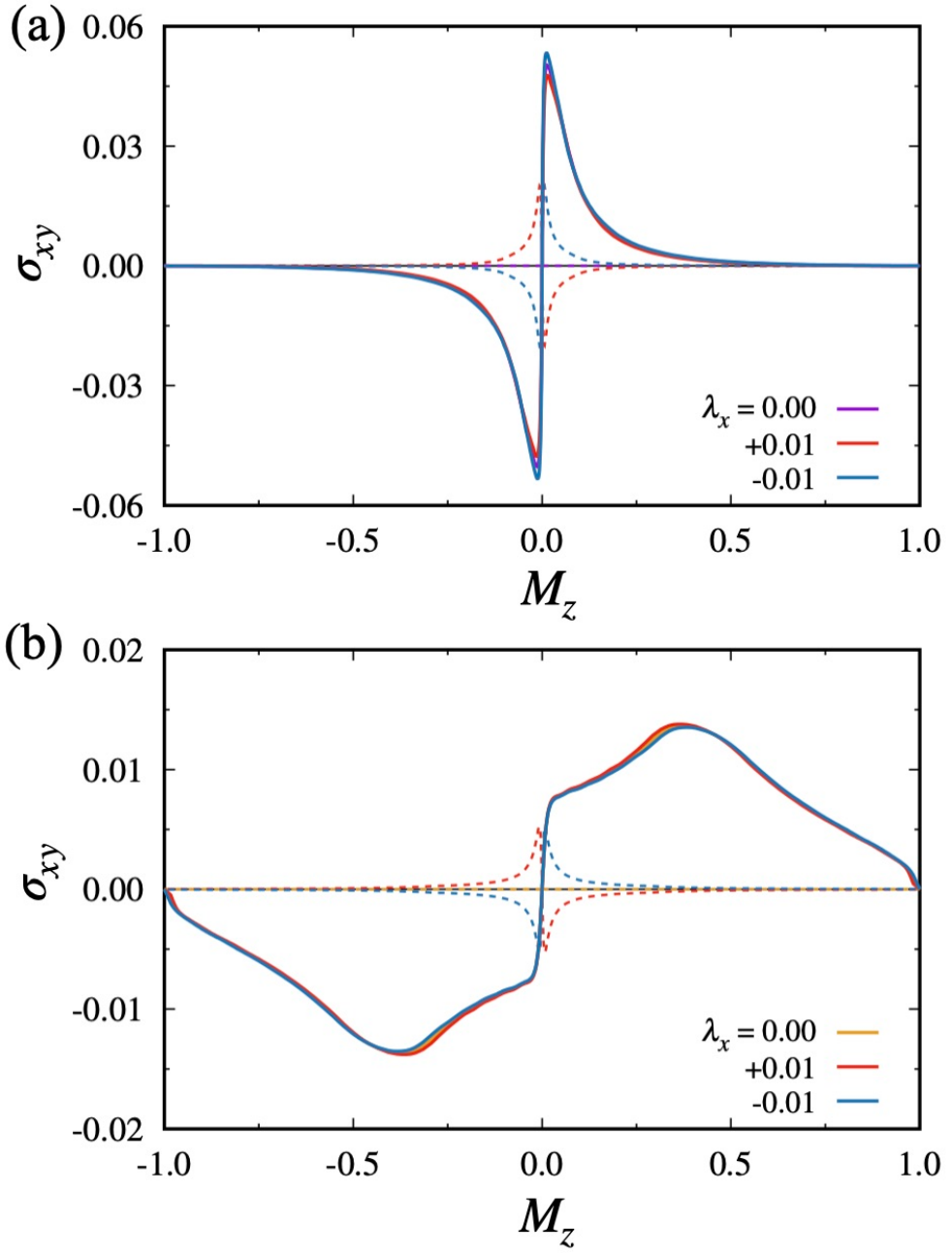}
\caption{
Net magnetization $M_z$ dependence of the AHC $\sigma_{xy}$ in the $yz$ spiral state for (a) $(J, n_\mathrm{e}) = (0.8, 0.005)$ and (b) $(2.0, 0.050)$ varying the $x$-bond SOC $\lambda_x$.
The dashed lines show the analogous plots in the ferromagnetic state.
}
\label{f6}
\end{figure}

Figure~\ref{f6} shows the AHC calculated with and without the $x$-bond SOC for the two representative parameter sets used in Fig.~\ref{f3}(a) of the main text: $(J, n_\mathrm{e}) = (0.8, 0.005)$ and $(2.0, 0.050)$, shown in Figs.~\ref{f6}(a) and \ref{f6}(b), respectively. 
In both cases, the magnetization dependence of the AHC in the spin spiral state remains almost unchanged for either sign of $\lambda_x$. 
The magnitude and sign are nearly identical to those obtained without the $x$-bond SOC.
This is in sharp contrast to the conventional AHE induced by SOC in the ferromagnetic state, where flipping only the sign of the $x$-bond SOC generally reverses the sign of the AHC, as indicated by the dashed lines. 

Furthermore, in comparison with the conventional AHE in the ferromagnetic state, the AHC in the spin spiral state is larger for the same parameter sets, especially for the larger magnetization.
Although the conventional AHE shows a nonmonotonic magnetization dependence similar to that of the SiAHE in Fig.~\ref{f6}(a), this originates from the general properties of a gapped two-dimensional Dirac point. 
As shown in Fig.~\ref{f6}(b), however, the magnetization dependence of the SiAHE significantly differs from the conventional AHE due to the anisotropic Fermi surface originating from odd-parity spin splitting with the spin spiral, as discussed in the main text.
The insensitivity of the SiAHE to $\lambda_x$ demonstrates that the characteristic Hall response is not governed by the detailed form of the SOC, but instead originates from the spiral-induced electronic structure.

This robustness can be understood in the low-energy effective model discussed in the main text. 
Including an additional spin--orbit term proportional to $k_x$, we can rewrite the two-level Hamiltonian as
\begin{eqnarray}
	H^{\mathrm{eff}}_{\bold{k}} = t(k_x^2 + k_y^2)\sigma^0 +(-p + \tilde{\lambda}_x)k_x\sigma^x + \tilde{\lambda}k_y\sigma^y -\tilde{m}\sigma^z,	\label{eq:lowE_model}	
\end{eqnarray}
where $\tilde{\lambda}_x$ is the reduced $x$-bond SOC in the continuum limit.
Depending on the relative sign of the SOC terms, the resulting low-energy Hamiltonian can be interpreted as realizing Dresselhaus-type ($\lambda_x < 0$) or Weyl-type ($\lambda_x > 0$) SOC.
Nevertheless, when the odd-parity band deformation $p$ is sufficiently large, the electronic structure is dominated by the spiral-induced spin splitting rather than by the $x$-bond SOC.
As a consequence, the additional $x$-bond SOC only weakly modifies the gapped spin-nodal lines and the associated Berry-curvature distribution responsible for the AHE. 
Its contribution to the AHC therefore remains negligible, justifying the simplified model adopted in the main text.

\subsection*{Odd-period spin spiral}

\begin{figure}[h]
\centering
\includegraphics[width=\columnwidth,clip]{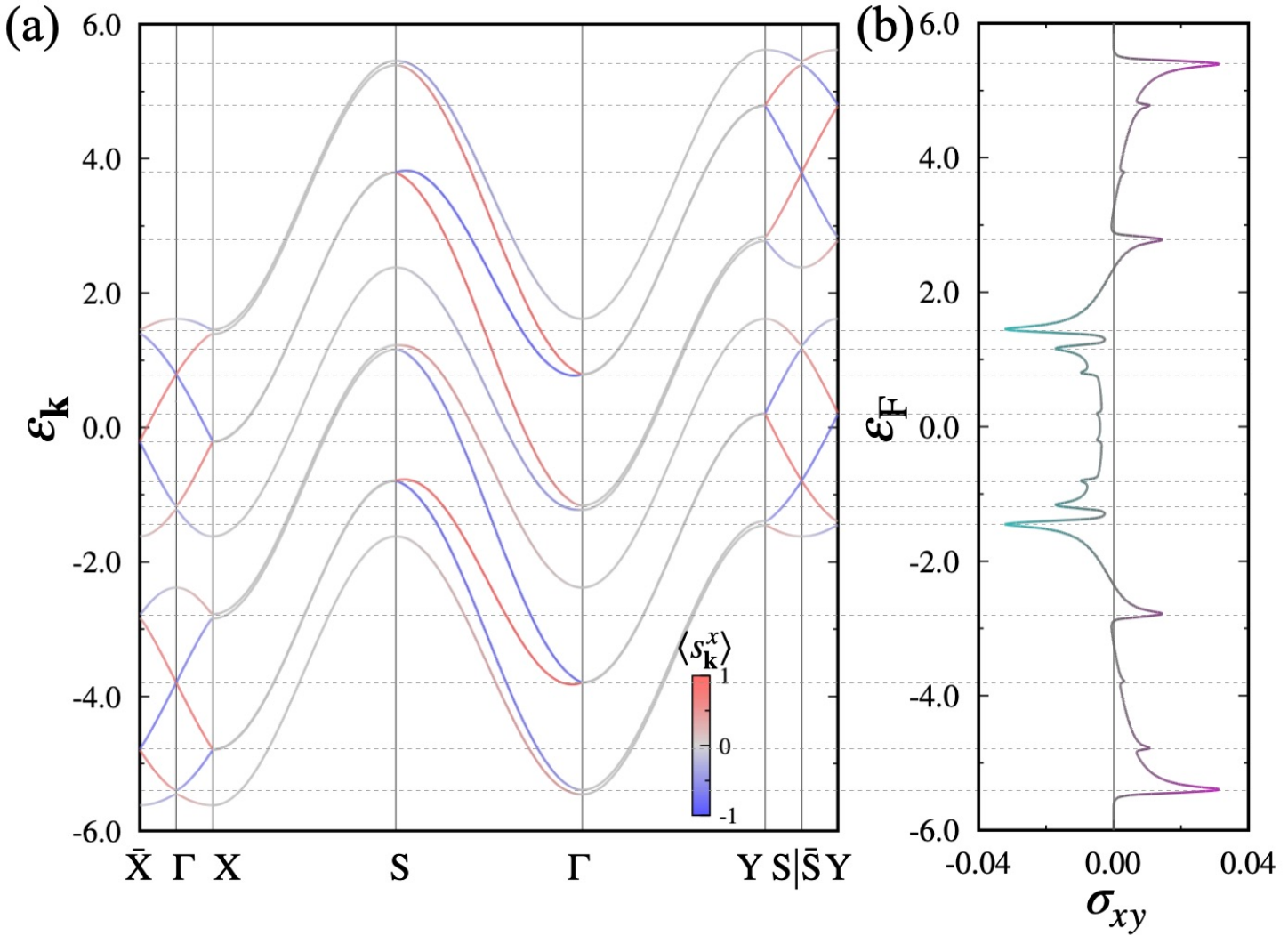}
\caption{
Similar plots to Fig.~\ref{f4} in the main text for $L = 5$ at $J = 2.0$, $\lambda = 0.01$, and $m = 0.1$.
}
\label{f7}
\end{figure}

We have focused on an even-period spiral with $L = 4$ in the main text. 
However, the phenomenon is universal for commensurate spirals, including those with odd periods.
We here discuss the case of the odd-period spin spiral with $L = 5$.
As shown in Fig.~\ref{f7}(a), spin splitting appears in an odd magnetic period as well, in essentially the same manner as in the even-period case.
Irrespective of the parity of $L$, $4(L-1)$ spin-nodal lines appear for a magnetic period $L$ along the $\Gamma$-Y and X-S lines at the zone boundaries of the folded BZ.
We confirm that these features are independent of the period and intrinsic to spiral magnetism protected by the symmetry of $C^{x}_{\pi}\mathcal{T}$ and $C^{x}_{Q} T^{\;}_{x}$, which are introduced in the main text.

Figure~\ref{f7}(b) shows the Fermi-energy $\varepsilon_\mathrm{F}$ dependence of the anomalous Hall conductivity $\sigma_{xy}$ for $L = 5$ at $J = 2.0$, $\lambda = 0.01$, and $m = 0.1$.
Reflecting the Berry curvature accumulated along the spin-nodal lines, the peak structures appear with an intensity comparable to that of the even-period case in Fig.~\ref{f4}(b) in the main text.
We therefore conclude that the spiral-induced anomalous Hall effect is a universal property of spin spirals.

\providecommand{\noopsort}[1]{}\providecommand{\singleletter}[1]{#1}%

\end{document}